\documentclass[lettersize,journal]{IEEEtran}
\usepackage{mathrsfs}
\usepackage{latexsym}
\usepackage{graphicx}
\usepackage{epsfig}
\usepackage{epstopdf}
\usepackage{subfigure}
\usepackage{array}
\usepackage{amsmath}
\usepackage{amssymb}
\usepackage{color, soul}
\usepackage{amsthm}
\usepackage{enumerate}
\usepackage{algpseudocode}
\usepackage{algorithm}
\usepackage{bm}
\usepackage{balance}

\theoremstyle{plain} 

\theoremstyle{definition}

\def\bal#1\eal{\begin{align}#1\end{align}}

\newcommand{\bW}{{\bf W}}
\newcommand{\bH}{{\bf H}}

\newcommand{\bn}{{\bf n}}

\newcommand{\bI}{{\bf I}}
\newcommand{\bU}{{\bf U}}

\newcommand{\bA}{{\bf A}}

\newcommand{\bQ}{{\bf Q}}

\newcommand{\bG}{{\bf G}}

\newcommand{\by}{{\bf y}}

\newcommand{\bx}{{\bf x}}

\newcommand{\bh}{{\bf h}}

\newcommand{\bq}{{\bf q}}

\newcommand{\bw}{{\bf w}}

\newcommand{\bp} {\begin{proof}}
	\newcommand{\ep} {\end{proof}}

\newcommand{{\Rb}} {\right)}

\newcommand{{\Rf}} {\right\}}
\begin{document}

\title{ \LARGE{RIS-Assisted Green Secure Communications: Active RIS or Passive RIS?}}

\author{Weigang Lv, Jiale Bai, Qingli Yan, Hui-Ming Wang, \emph{Senior Member, IEEE}

\thanks{The work was supported in part by the National Natural Science Foundation of China under Grants 61941105 and 62171364.}

\thanks{Weigang Lv, Jiale Bai and Hui-Ming Wang are with the School of Information and Communications Engineering, Xi’an Jiaotong University, Xi’an 710049, China (\emph{Corresponding author: Hui-Ming Wang}.)


Qingli Yan is with the School of Computer Science $\&$ Technology, Xi’an University of Posts $\&$ Telecommunications, Xi’an 710121, China (e-mail: yql@xupt.edu.cnm.)}}

\maketitle

 \pagenumbering{arabic}

\begin{abstract}
Reconfigurable Intelligent Surface (RIS) is one of the promising techniques for 6G wireless communications, and recently has also been shown to be able to improve secure communications. However, there is a “double fading” effect in the reflection link between base station and user, thus passive RIS only achieves a negligible secrecy gain in typical communications scenarios. In this letter, we propose an active RIS-aided multi-antenna physical layer secrecy transmission scheme, where  the active RIS can amplify the signal actively. Our aim is to minimize the transmit power subject to the constraint of secrecy rate. To solve the non-convex optimization problem, a penalty-based alternating minimization (AltMin) algorithm is proposed to optimize both the beamformer at the transmitter and the reflection matrix at RIS. Simulation results show that active RIS can resist the impact of “double fading” effect effectively, and is more energy efficient than passive RIS.
\end{abstract}


\section{Introduction}

With the development of the fifth-generation mobile communication technology (5G), network energy consumption becomes a growing concern. At the same time, a revolutionary technique called RIS has been proposed. RIS, a planar array, is usually composed of a large number of low-cost, passive, reflecting units, and each unit can reflect the incident wireless signal with an adjustable phase shift. Due to the low energy consumption characteristics of RIS, passive RIS-aided system has received considerable attention \cite{Wu-02}, \cite{Shen-19}. On the other hand, the security of communications is also an important issue. Traditionally, a secure transmission is realized by cryptographic methods, and in recent years, the concept of passive RIS-aided physical layer security has been proposed \cite{Wu-03}, \cite{Wu-04}.

At present, secure communications with low power consumption is one of the key points in RIS research. However, recent studies show that the reflected signals of RIS go through two paths: base station-RIS and RIS-User paths, and the signals received by users will be affected by the “double fading” effect, thus the performance gain of RIS is limited. In order to alleviate the impact of “double fading”, the concept of active RIS is proposed in \cite{Wu-05}, \cite{Wu-06}. Different from the traditional passive RIS, the feature of active RIS is that each element of RIS can adjust the phase of the signal and amplify the signal simultaneously. In addition, it is pointed out in \cite{Wu-07} that active RIS is fundamentally different from conventional relay.

All the existing works for the study of green secure RIS assisted system have ignored an unavoidable problem: although RIS brings new reliable reflection link for signal transmission in addition to the direct link, a negative “double fading” effect always exists in this link, i.e., the signals received via this link suffer from large-scale fading twice, resulting in a limited performance gain compared with the one without RIS. 
On the other hand, some research results over active RIS assisted system have been established. However, although active RIS greatly helps enhancing the quality of communications of user, it also reduces the “ double fading ” effect in the reflection link of base station-RIS-eavesdropper due to the broadcast nature of wireless channels, resulting larger information leakage to eavesdropper. Motivated by the above discussions, in this letter, we consider an active RIS-aided multi-antenna secure wireless communications system, in which a multi-antenna base station (Alice) communicates with a single-antenna user through active RIS in the presence of a single-antenna eavesdropper. The main contributions of this letter are summarized as follows:

$\bullet$ First, to the best of our knowledge, this is the first work that employs active RIS help reduce the secure transmission power consumption at the transmitter in multi-antenna wiretap channels. We compare the active RIS with the conventional passive RIS, by formulating the respective power minimization of the transmitter problems under the same secrecy rate constraint.
	
$\bullet$ Second, we solve the non-convex $P1$ via an iterative algorithm which alternatively optimizes the beamforming $\mathbf{w}$ and the reflecting coefficient matrix $\mathbf{Q}$ at the RIS. In particular, we employ penalty method to deal with the nonconvex rank-1 constraint, and optimal $\mathbf{w},\ \mathbf{Q}$ are obtained via successive convex approximation (SCA) algorithm.
	
$\bullet$ Finally, simulation results are provided to demonstrate that under the setting of the same secrecy rate, employing active RIS to assist secure transmission consumes less energy than passive RIS, which indicates that active RIS-aided scheme is effective on weakening the influence caused by “double fading” effect.

\emph{Notations}: 
$\ \mathbb{C}^{m\times n}$ stands for the set of all $m\times n$ complex valued matrices; $\mathbb{H}^{m}$ represents the set of all $m\times m$ Hermitian matrices; $\bA=diag(\mathbf{y})$ represents $\bA$ is a diagonal matrix composed of elements in $\mathbf{y}$, where $\mathbf{y}$ is a vector; the largest eigenvalue of matrix $\mathbf{X}$ and its associated eigenvector are denoted by $\lambda_{max}(\mathbf{X})$ and $\pmb{\nu}_{max}(\mathbf{X})$; $rank(\mathbf{X}),\ Tr(\mathbf{X})$ denotes the rank and the trace of matrix $\mathbf{X};\ \mathbf{X}\succeq0$ indicates that $\mathbf{X}$ is a positive semidefinite matrix; $\by[1:n]$ is to extract the first $n$ entires in $\by$; $\nabla F$ denotes the gradient of a function $F$; $dom F$ denotes the domain of a function $F$.

\section{Channel Model And Problem Formulation}
We consider an active RIS-aided multiple-input single-output (MISO) wiretap communications system as in Fig. 1. The system consists of base station (Alice), user (Bob) and eavesdropper (Eve), as well as an active RIS. In this model active RIS consists of $N$ reflecting elements, Alice is equipped with $M$ antennas, both Bob and Eve are equipped with a single antenna. The channel coefficients of the Alice-Bob link, the Alice-Eve link, the Alice-RIS link, the RIS-Bob link, and the RIS-Eve link are denoted by $\bh_{AB}\in\mathbb{C}^{M\times 1}, \bh_{AE}\in\mathbb{C}^{M\times 1}, \bH_{AI}\in\mathbb{C}^{N\times M}, \bh_{IB}\in\mathbb{C}^{N\times 1}$, and  $\bh_{IE}\in\mathbb{C}^{N\times 1}$, respectively. We donate $\bQ \triangleq \emph{diag}(\beta_1e^{j\theta_1},\beta_2e^{j\theta_2},\cdots,\beta_ne^{j\theta_N})$ as the diagonal matrix associated with the phase shifts and reflecting coefficient in all RIS elements, where $\theta_n\in[0,2\pi]$ and $\beta_n\ge 0$ is the phase shift and the amplification factor at the $n$-th element, respectively. Note that, in active RIS it is possible that $\beta_n >1$. 

The received signal at Bob and Eve can be written, respectively, as
\begin{align}
y_b = (\bh^{H}_{AB}+\bh^{H}_{IB}\bQ\bH_{AI})\bx+\bh^{H}_{IB}\bQ\bn_I+n_b, \notag\\y_e = (\bh^{H}_{AE}+\bh^{H}_{IE}\bQ\bH_{AI})\bx+\bh^{H}_{IE}\bQ\bn_I+n_e, \notag
\end{align}
where $\bx=\bw s$ is the transmitted signal, $s$ denotes the desired signal for Bob with $\mathbb{E}\left\{|s|^2 \right\}=1$, and $\bw\in\mathbb{C}^{M\times 1}$ is the transmit beamformer. $n_b\sim\mathcal{CN}(0,\sigma_B^{2})$, and $n_e\sim\mathcal{CN}(0,\sigma_E^{2})$ are additive complex white Gaussian noises at Bob and Eve respectively. Note that, different from passive RIS, active RIS requires extra power, and due to each element is equipped with an amplifier, we cannot ignore the thermal noise $\bn_I\sim\mathcal{CN}(\mathbf{0},\sigma ^2_I\bI)$ generated at RIS.

To investigate the potential maximum improvement on secrecy by active RIS,  
similar to \cite{Wu-08}, we assume that instantaneous channel state information (CSI) of all the channel links are available at Alice, which provides a benchmark for all possible practical schemes. Accordingly, the achievable rates of the legitimate and the eavesdropper links can be given by   
\begin{align}
R_B\triangleq\ln\left(1+\frac{|(\bh^{H}_{AB}+\bh^{H}_{IB}\bQ\bH_{AI})\bw|^2}{{\sigma_B}^2+||\bh^{H}_{IB}\bQ||^2{\sigma_I}^2}\right), \notag \\ R_E\triangleq\ln\left(1+\frac{|(\bh^{H}_{AE}+\bh^{H}_{IE}\bQ\bH_{AI})\bw|^2}{{\sigma_E}^2+||\bh^{H}_{IE}\bQ||^2{\sigma_I}^2}\right). \notag
\end{align}
The achievable secrecy rate is given by $R_{s}=[R_B-R_E]^{+}$, where $[x]^{+}=max(x,0)$. In this letter, we are interested in whether active RIS-aided scheme is better than passive RIS-aided scheme in improving the energy efficiency of secure communications, so we aim to minimize the total transmit power at the BS subject to the secrecy rate constraints. Then the proposed power efficient design of the beamformer at the Alice and the reflecting elements at the RIS are obtained by solving the following optimization problem:
\begin{align}
P1: \underset{\bw, \bQ}{\min}&\quad ||\bw||^{2} \notag \\ 
s.t.&\quad R_{s}\geq \bar{R}\tag{1a},\\ &\quad ||\bQ\bH_{AI}\bw||^2+||\bQ||_{F}^2\sigma_I^{2}\leq P_{I},\tag{1b} 
\end{align}
where $\bar{R}$ is the target secrecy rate, $P_{I}$ is the maximum amplification power budget at RIS. 
Problem $P1$ is non-convex due to the complicated form of secrecy rate constraint (1a), thus it cannot be solved directly to jointly design the optimal $\bw$ and $\bQ$. Next, we propose an alternating optimization algorithm to solve this problem.
\begin{figure}[t]
	\centerline
	{\includegraphics[width=2.5in]{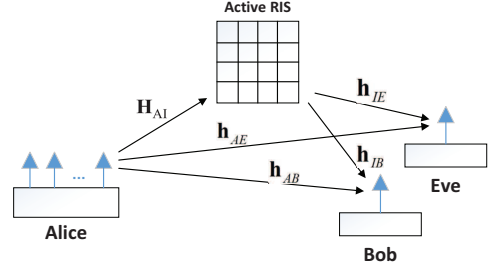}}
	\caption{An active RIS-assisted MISO communications system}
\end{figure}

\section{Alternating Minimization Algorithm}
In this section, we propose a penalty-based AltMin algorithm to design the transmit beamformer $\bw$ and $\bQ$ alternatively. Specifically, $\bw$ and $\bQ$ are optimized in two sub-problems. When $\bQ$ is fixed, the problem becomes a semi-definite relaxation (SDR) problem where 
the rank-one constraint is usually recovered by Gaussian randomization. We employ an equivalent representation of the rank-one constraint, and add a penalized version of the equivalent representation to the objective function, then SCA algorithm is used to solve this optimization problem. When $\bw$ is fixed, we set a variable $\delta$ to avoid the feasibility check problem, and then we apply the same penalty-based method in combination with SCA algorithm to obtain the optimal solution of $\bQ$. Finally, we can obtain a Karush-Kuhn-Tucker (KKT) solution of $P1$.

\subsection{Optimal Secure Transmit Beamformer \bw}
When $\bQ$ is fixed, $P1$ can be reduced to the following sub-problem $P2$.
\begin{align}
P2: \underset{\bw}{\min}&\quad ||\bw||^{2}\notag \\ 
\quad s.t.&\quad \ln\left(1+\frac{|(\bh^{H}_{AB}+\bh^{H}_{IB}\bQ\bH_{AI})\bw|^2}{{\sigma_B}^2+||\bh^{H}_{IB}\bQ||^2{\sigma_I}^2}\right)-\notag\\&\qquad\ln\left(1+\frac{|(\bh^{H}_{AE}+\bh^{H}_{IE}\bQ\bH_{AI})\bw|^2}{{\sigma_E}^2+||\bh^{H}_{IE}\bQ||^2{\sigma_I}^2}\right) \geq \bar{R},\tag{2a}\\&\quad ||\bQ\bH_{AI}\bw||^2+||\bQ||_{F}^2\sigma_I^{2}\leq P_{I}.\tag{2b}  
\end{align}

Due to constant (2a), problem $P2$ is non-convex. We denote $\bh_{B}=(\bh^{H}_{AB}+\bh^{H}_{IB}\bQ\bH_{AI})^{H},\  \bh_{E}=(\bh^{H}_{AE}+\bh^{H}_{IE}\bQ\bH_{AI})^{H}$, and $\bW=\bw\bw^{H}$. We have 
\begin{align}
P3: \underset{\bW}{\min}&\quad tr(\bW)\notag \\ 
s.t.&\quad\dfrac{1}{{\sigma_B}^2+||\bh_{IB}\bQ||^2{\sigma_I}^2}tr(\bh_{B}\bh_{B}^{H}\bW)-\notag\\&\qquad \dfrac{1}{{\sigma_E}^2+||\bh_{IE}\bQ||^2{\sigma_I}^2}tr(\bh_{E}\bh_{E}^{H}\bW) \geq e^{\bar{R}}-1,\tag{3a}\\&\quad tr(\bQ\bH_{AI}\bW\bH_{AI}^{H}\bQ^{H})\leq P_{I}-||\bQ||_{F}^2\sigma_I^{2}, \tag{3b} \\&\quad\bW \succeq 0,\tag{3c}
\end{align}
where the non-convex $rank(\bW)=1$ constraint is omitted. Problem $P3$ is a standard SDP, which can be directly optimized. However, the solution may not satisfy the rank-one constraint. To
recover the optimal rank-one solution, we change the rank-one constraint into an alternative representation in the following lemma.

\emph{Lemma 1}: The rank-one constraint is equivalent to ${(3d)}$.
\begin{align}
tr(\bW)-\lambda_{max}(\mathbf{W})\leq 0 \tag{${3d}$}.
\end{align}
\begin{proof}
For any $\bA\in\mathbb{H}^{m}$, the inequality $tr(\bA)=\Sigma_{i}\rho_{i} \geq \lambda_{max}(\mathbf{A}) = max{(\rho_{i})}$ holds, where $\rho_{i}$ is the $i$-th eigenvalue value of $\bA$. Equality holds if and only if $\bA$ has unit rank.
\end{proof}
Then, we propose a penalty-based method \cite{Wu-09} to tackle the rank-one constraint. Specifically, by employing a penalty factor to the constraint ${(3d)}$ and then adding it to the objective function, problem $P3$ is transformed to
\begin{align}
P4: \underset{\bW}{\min}&\quad tr(\bW)+\dfrac{1}{\eta}(tr(\bW)-\lambda_{max}(\mathbf{W})) \notag\\ s.t.&\quad \dfrac{1}{{\sigma_B}^2+||\bh^{H}_{IB}\bQ||^2{\sigma_I}^2}tr(\bh_{B}\bh_{B}^{H}\bW)-\notag\\&\qquad\dfrac{1}{{\sigma_E}^2+||\bh^{H}_{IE}\bQ||^2{\sigma_I}^2}tr(\bh_{E}\bh_{E}^{H}\bW) \geq e^{\bar{R}}-1,\tag{4a}\\&\quad (3b),(3c),\notag
\end{align}

where $\eta\textgreater0$ is the penalty factor. According to the key idea of proof in \cite{Wu-09}, When $\eta$ is sufficiently small, $P3$ and $P4$ have the same optimal solution. We propose SCA algorithm to solve this problem. The key idea of SCA algorithm is obtain an approximately surrogate function and iteratively optimize the solution. Specifically, we construct a surrogate function of the objective function $P4$ by employing the following first-order Taylor approximation:

\begin{align}
\lambda_{max}(\mathbf{W})\geq& \lambda_{max}(\bW^{(t)})+\notag\\ &tr[\pmb{\nu}_{max}(\bW^{(t)})\pmb{\nu}_{max}^{H}(\bW^{(t)})(\bW-\bW^{(t)})], \notag
\end{align}
where $\bW^{(t)}$ is the solution of the $t$-th iteration. Finally, the optimization problem $P4$ can be expressed in the following form:
\begin{align}
P5: \underset{\bW}{\min}\quad &tr(\bW)+\dfrac{1}{\eta}(tr(\bW)-\lambda_{max}(\bW^{(t)})-\notag\\ &\quad tr[\pmb{\nu}_{max}(\bW^{(t)})\pmb{\nu}_{max}^{H}(\bW^{(t)})(\bW-\bW^{(t)})]) \notag\\ s.t.\quad& \dfrac{1}{{\sigma_B}^2+||\bh^{H}_{IB}\bQ||^2{\sigma_I}^2}tr(\bh_{B}\bh_{B}^{H}\bW)-\notag\\&\quad \dfrac{1}{{\sigma_E}^2+||\bh^{H}_{IE}\bQ||^2{\sigma_I}^2}tr(\bh_{E}\bh_{E}^{H}\bW) \geq e^{\bar{R}}-1,\tag{5a}\\&\quad (3b),(3c),\notag
\end{align}
which is a convex optimization problem and therefore can be solved iteratively. The penalty-based algorithm for recovering rank-one is summarized in Algorithm 1.
\begin{algorithm}[t]
	\caption{(Penalty-based SCA algorithm)}
	\label{alg.2}
	\begin{algorithmic}
		\State 1.Initialize  with a value of the feasible point. Set the convergence tolerance $0 \textless \varepsilon_1 \ll 1$, penalty factor $0 \textless \eta \ll 1 $ and iteration index $t=0$.
		\Repeat\ \ 
		\State 2. Calculate $\bW^{(t)}$ in $P5$ with given feasible point.
		\State 3. if $|tr(\bW^{(t)})-\lambda_{max}(\bW^{(t)})| \textless \varepsilon_1 $, stop algorithm and go to step 6, else go to step 4.
		\State 4. For given $\bW^{(t)}$ and given \bQ, update $\bW^{(t+1)}$ in $P5$.
		\State 5. $t=t+1,\bW^{(t)}=\bW^{(t+1)}$ go to step 3.
		\Until $|tr(\bW^{(t)})-\lambda_{max}(\bW^{(t)})| \textless \varepsilon_1 $
		\State 6.Output $\bW^{(t)}$ as the optimal rank-1 solution for $P3$.
	\end{algorithmic}
\end{algorithm}
 Then the optimal beamformer of $P3$ can be expressed as
 $\bw=\pmb{\nu}_{max}(\bW)\sqrt{\lambda_{max}(\bW)}$. Note that, by applying SCA, the minimum objective value of problem $P5$ serves as an upper bound for the optimal value of problem $P4$. By iteratively solving problem $P5$ optimally, we can monotonically tighten this upper bound. In this way, the objective values achieved by the sequence $ \{ \bW^{(t)} \} $ form a non-increasing sequence that converges to a stationary point of problem $P4$ in polynomial time.

\subsection{Optimizing \bQ\ with given \bw}
In this subsection, we design $\bQ$ at RIS for given a $\bw$. We denote $\bq=[ \beta_1e^{j\theta_1},\beta_2e^{j\theta_2},\cdots,\beta_ne^{j\theta_N}]^{H},\  \bG_{i}=diag(\bh^{H}_{Ii})\bH_{AI}$, and the sub-problem of optimizing $\bQ$ is equivalently modified as
\begin{align}
P6:find&\quad \bq \notag \\ s.t. &\quad \widehat{R_{B}}-\widehat{R_{E}}\geq \bar{R}, \tag{6a}\\&\quad ||\bq^{H}diag{(\bH_{AI}\bw)}||^2+||\bq||^2\sigma_I^{2}\leq P_{I}, \tag{6b} 
\end{align}
where $\ \forall i \in \{B,E\} ,\ \widehat{R_{i}}=\ln(\sigma_i^{2}+||\bq^{H}diag{(\bh^{H}_{Ii})}||^2 \sigma_I^{2}+|(\bh^{H}_{Ai}+\bq^{H}\bG_{i})\bw|^2)- \ln(\sigma_i^{2}+||\bq^{H}diag{(\bh^{H}_{Ii})}||^2 \sigma_I^{2}) $. 

$P6$ is a feasibility check problem. In order to improve the converged solution in the optimization process, we set a variable $\delta$ to avoid this feasibility check problem \cite{Wu-10}, then problem $P6$ is expressed as
\begin{align}
P7:\underset{\bq,\delta}{\max}&\quad \delta \notag \\ s.t. &\quad \widehat{R_{B}}-\widehat{R_{E}}-\delta\geq \bar{R},\ \delta \geq 0, \tag{7a}\\&\quad||\bq^{H}diag{(\bH_{AI}\bw)}||^2+||\bq||^2\sigma_I^{2}\leq P_{I}. \tag{7b}
\end{align}

In order to solve $P7$, we set $\bU=[\bq^{H}\quad1]^{H}[\bq^{H}\quad1]$ with $rank(\bU)=1$, and the same penalty-based algorithm in sub-problem 1 is used to deal with the rank-one constraint. Now $P7$ can be expressed as
\begin{align}
P8: \underset{\bU,\delta}{\max}&\quad \delta-\dfrac{1}{\eta}(tr(\bU)-\lambda_{max}(\bU^{(t)})-\notag\\ &\qquad tr[\pmb{\nu}_{max}(\bU^{(t)})\pmb{\nu}_{max}^{H}(\bU^{(t)})(\bU-\bU^{(t)})]) \notag \\ s.t.&\quad \widetilde{R_{B}}-\widetilde{R_{E}}-\delta\geq \bar{R},\ \delta \geq 0, \tag{8a} \\&\quad tr(\bG\bU) \leq P_{I},\tag{8b} \\&\quad\bU \succeq 0, \quad \bU[N+1,N+1]=1, \tag{8c} 
\end{align}
where $\bU^{(t)}$ is the solution of the $t$-th iteration, $ \widetilde{R_{i}}=\ln(tr(\bG_{Ai}\bU))-\ln(tr(\bG_{Ii}\bU))$, 

\begin{align}
 &\bG_{Ii}=\left[\begin{array}{cc}
\sigma_I^{2}diag(\bh^{H}_{Ii})diag(\bh^{H}_{Ii})^{H} & \bm{0}_{n\times 1}\\
\bm{0}_{1\times n}& \sigma_i^{2}
\end{array} 
\right], \notag\\
&\bG_{Ai}=\left[\begin{array}{cc}
\bG_{i}\bw\bw^{H}\bG_{i}^{H}+\bar{\bG}_{i} & \bG_{i}\bw\bw^{H}\bh_{Ai}\\
\bh_{Ai}^{H}\bw\bw^{H}\bG_{i}^{H}& \mu_{i}
\end{array} 
\right], \notag\\
&\bG=\left[\begin{array}{cc}
diag(\bH_{AI}\bw)diag(\bH_{AI}\bw)^{H}+\sigma_I^{2}\bI & \bm{0}_{n\times 1}\\
\bm{0}_{1\times n}& 0
\end{array} 
\right], \notag
\end{align}
and $\mu_{i}=\sigma_i^{2}+\bh^{H}_{Ai}\bw\bw^{H}\bh_{Ai},\  \bar{\bG}_{i}=\sigma_I^{2}diag(\bh^{H}_{Ii})diag(\bh^{H}_{Ii})^{H}$.

Problem $P8$ is still intractable due to the non-convex constraint $(8a)$, we propose SCA algorithm to solve this problem as well, where the surrogate function of the constraint (8a) is shown in the following lemma. 

\emph{Lemma 2}: for given feasible point $\bU^{(t)}$, the following inequalitie always holds 
\begin{align}
\hat{R}(\bU)=&\widetilde{R_{B}}-\widetilde{R_{E}} \geq \notag \\&\ln(tr(\bG_{AB}\bU))+\ln(tr(\bG_{IE}\bU))\notag \\&-\ln(tr(\bG_{IB}\bU^{(t)}))-\ln(tr(\bG_{AE}\bU^{(t)}))\notag \\&-tr\left(\dfrac{\bG_{IB}}{tr(\bG_{IB}\bU^{(t)}) }(\bU-\bU^{(t)}) \right)\notag \\&-tr\left(\dfrac{\bG_{AE}}{tr(\bG_{AE}\bU^{(t)})}(\bU-\bU^{(t)}) \right)\notag\\& = \widetilde{R_{S}}(\bU;\bU^{(t)}).  \notag
\end{align}
\begin{proof}
By applying the key idea of first-order Taylor expansion theorem \cite{Wu-11}:

For any differentiable function $F(x)$ and $\bar{x}\in dom F$, it always holds that $F(x) \geq F(\bar{x})+(\nabla F(\bar{x}))^{T}(x-\bar{x})$ if $F(x)$ is convex, and $F(x) \leq F(\bar{x})+(\nabla F(\bar{x}))^{T}(x-\bar{x})$ if $F(x)$ is concave. Due to $\ln(tr(\bG_{IB}\bU))$ and $  \ln(tr(\bG_{AE}\bU))$ are both concave respect to $\bU$. Then the following two inequalities always holds
\begin{align}
\ln(tr(\bG_{IB}\bU)) \leq&\notag \ln(tr(\bG_{IB}\bU^{(t)}))\\&+tr\left(\dfrac{\bG_{IB}}{tr(\bG_{IB}\bU^{(t)}) }(\bU-\bU^{(t)}) \right), \notag\\\ln(tr(\bG_{AE}\bU)) \leq&\notag \ln(tr(\bG_{AE}\bU^{(t)}))\\&+tr\left(\dfrac{\bG_{AE}}{tr(\bG_{AE}\bU^{(t)}) }(\bU-\bU^{(t)}) \right).\notag
\end{align}

\end{proof}
According to lemma 2, $P8$ can be approximated as
\begin{align}
P9: \underset{\bU,\delta}{\max}&\quad \delta-\dfrac{1}{\eta}(tr(\bU)-\lambda_{max}(\bU^{(t)})-\notag\\ &\qquad tr[\pmb{\nu}_{max}(\bU^{(t)})\pmb{\nu}_{max}^{H}(\bU^{(t)})(\bU-\bU^{(t)})]) \notag \\ s.t.&\quad \widetilde{R_{S}}(\bU;\bU^{(t)})-\delta \geq \bar{R},\ \delta \geq 0, \tag{9a} \\&\quad (8b),(8c). \notag
\end{align}

The problem $P9$ is convex, so it can be directly solved. Finally, after obtaining the optimal $\bU,\ \bQ$ can be reformulated as $\bQ=diag( (\pmb{\nu}_{max}(\bU)\sqrt{\lambda_{max}(\bU)})[1:N])$. Then the penalty-based SCA algorithm for optimize $\bU$ is summarized in Algorithm 2.

\subsection{Overall Algorithm}
Denote $g(\mathbf{w},\mathbf{Q})$ as the objective function of $P1$, then consider that $\mathbf{w}^{(t)}$ is the solution of $P5$, and $\mathbf{Q}^{(t)}$ is the solution of $P9$, one obtains that $g(\mathbf{w}^{(t+1)},\mathbf{Q}^{(t+1)}) \leq g(\mathbf{w}^{(t)},\mathbf{Q}^{(t)})$, and $g(\mathbf{w},\mathbf{Q})\geq0$. Thus, the AltMin algorithm is guaranteed to have a monotonic convergence. In the AltMin algorithm, the main computational complexity for optimizing $\mathbf{w}$ given $\mathbf{Q}$ and optimizing $\mathbf{Q}$ given $\mathbf{w}$ are about $\mathcal{O}(M^{3})$ and $\mathcal{O}((N + 1)^{3})$ respectively.
\begin{algorithm}[t]
	\caption{(Penalty-based SCA algorithm )}
	\label{alg.2}
	\begin{algorithmic}
		\State 1.Initialize $\bU^{(0)}$ with a value of the feasible point. Set the convergence tolerance $0 \textless \varepsilon_2 \ll 1$ and iteration index $t=0$, according to $\bU^{(0)}$, compute $tr(\bG\bU^{(t)})$  
		\Repeat\ \ 
		\State 2. Calculate $\bU^{(t+1)}$ in $P9$ with given $\bU^{(t)}$
		\State 3. According to $\bU^{(t+1)}$, compute $tr(\bG\bU^{(t+1)})$
		\State 4. if $|tr(\bG\bU^{(t+1)})-tr(\bG\bU^{(t)})| \leq \varepsilon_2$, stop algorithm 2, else $t=t+1$ go to step 2.
		\Until $|tr(\bG\bU^{(t+1)})-tr(\bG\bU^{(t)})| \leq \varepsilon_2$
		\State 5.Output $\bU$ as the optimal rank-one solution for $P9$.
	\end{algorithmic}
\end{algorithm}

\section{Simulation Results}
Numerical simulations have been carried out in this section. Following \cite{Wu-04}, we consider that all the channels are formulated as the product of large scale fading and small scale fading. Since RIS is usually deployed on high places with line-of-sight (LOS) path, small scale fading in all channels are assumed to be Rician fading.
 
We set the location of each node as $l_{A}=(0,0),\ l_{I}=(60,30),\ l_{B}=(70,20),\ l_{E}=(100,20)$, and noise power $\sigma^2_B=\sigma^2_E=\sigma^2_I=-90$dBm. In the AltMin algorithm, all the target accuracy is set as $10^{-3}$, and the starting point is set as $\bw=0,\ \bQ=I$. All the results are averaged over 100 channel realizations.
\begin{figure}[htbp]
	\centering
	{\includegraphics[scale=0.4]{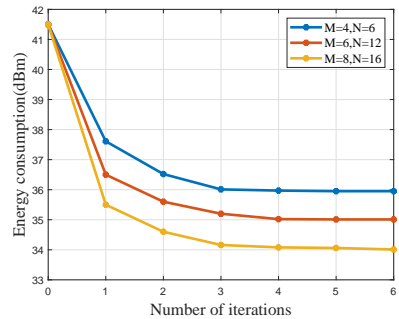}}
	\caption{ Convergence of proposed algorithms under $\bar{R}=10$.}
	\label{1}
\end{figure}

First, the convergence of penalty-based AltMin algorithm for optimizing Q and w is shown in Fig. 2. Note that the total energy consumption with the active RIS is calculated as $P_{total}=P_{t}+||\bQ\bH_{AI}\bw||^2+||\bQ||^2_{F}\sigma^2_I$, in which $P_{t}$ is transmit power. This figure verifies the convergence and effectiveness of the proposed AltMin algorithm.
\begin{figure}[htbp]
	\centering
	{\includegraphics[scale=0.4]{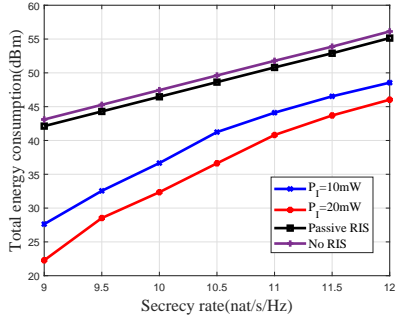}}
	\caption{ Total energy consumption versus secrecy rate.}
	\label{1}
\end{figure}

Then, we can observe from Fig. 3 that due to the “double fading” effect, passive RIS-aided scheme can only reduces a small part of the transmit power. However, under the setting of the same secrecy rate, employing active RIS consumes less energy than passive RIS, which indicates that active RIS-aided scheme is effective on weakening the influence caused by “double fading” effect. And we found that although active RIS requires extra power consumption, only a small part of the total power consumption is used for signal amplification. Moreover, increasing the maximum power allowance at the RIS from 10 mW to 20 mW leads to further transmit power savings. Hence, the active RIS-aided scheme achieves a better energy efficiency than passive-aided scheme. 
\begin{figure}[htbp]
	\centering
	{\includegraphics[scale=0.4]{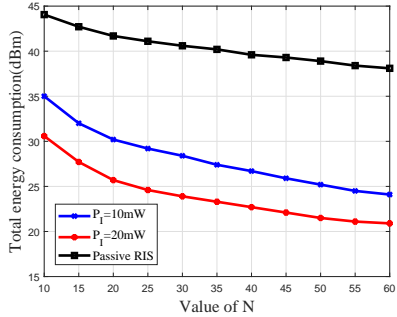}}
	\caption{Total energy consumption versus reflecting surface elements N.}
	\label{1}
\end{figure}

In addition, Fig. 4 shows the total energy consumption versus the number of reflecting surface elements $N$ with given $\bar{R}=10$. From Fig. 4, it is observed that as the number of reflecting elements increases, the total energy consumption required by the system decreases. Furthermore, note that due to the “double fading” effect, the energy consumption with passive RIS only reduces by about 20$\%$ when $N$ varies from 10 to 60, which is less than 35$\%$ achieved by active RIS when $P_{I}$=10mW. The result fully indicates that using active RIS can help us save the number of reflecting elements to achieve a better performance gain compared with passive RIS case.

Finally, to explore the impact of location on RIS performance, we set the location of each node as $l_{A}=(0,0),\ l_{I}=(d_{s},0),\ l_{B}=(70,0),\ l_{E}=(100,0)$, and the RIS is movable on the $x$-axis. Fig. 5 shows that as $d_{s}$ increases, the total energy consumption decreases, i.e., when the RIS is placed near Bob, RIS can achieve a better performance gain. In addition, due to the “double fading” effect, the secure transmission power of active RIS-aided scheme is always lower than passive RIS-aided scheme. Note that, passvie RIS is actually a special case of active RIS. When we set the reflection coefficient of active RIS to 1, active RIS is the same as passive RIS.

\begin{figure}[htbp]
	\centering
	{\includegraphics[scale=0.4]{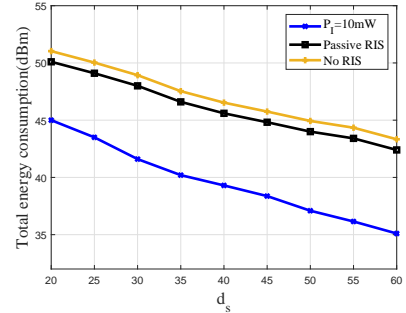}}
	\caption{Total energy consumption versus $d_{s}$, M=4 and N=8 , $\bar{R}=10$.}
	\label{1}
\end{figure}


\section{Conclusion}
In this letter, we proposed an active RIS-aided multi-antenna physical layer secrecy transmission scheme. We investigated to minimize the transmit power subject to the secrecy rate constraint and the simulation results show that Active RIS-aided scheme achieves a better energy efficiency. On the other hand, active RIS can reduce the number of reflecting units, thus reducing the RIS area and making it easier to deploy in practical applications.


\begin{thebibliography}{99}

\bibitem{Wu-02}C.-W. Huang, et al., ``Holographic MIMO surfaces for 6G wireless networks: opportunities, challenges, and trends," \emph{IEEE Wireless Commun.}, vol. 27, no. 5, pp. 118-125, Oct. 2020.

\bibitem {Shen-19}M. Di Renzo, et al., ``Smart radio environments empowered by reconfigurable intelligent surfaces: How it works, state of research, and the road ahead," \emph{IEEE J. Sel. Areas Commun.}, vol. 38, no. 11, pp. 2450-2525, Nov. 2020.


\bibitem{Wu-03} L.-M. Dong, H.-M. Wang, ``Secure MIMO transmission via intelligent reflecting surface," \emph{IEEE Wirel. Commun. Lett.}, vol. 9, no. 6, pp. 787-790, Jun. 2020.


\bibitem{Wu-04} M. Cui, G. Zhang, R. Zhang, ``Secure wireless communication via intelligent reflecting surface," \emph{IEEE Wirel. Commun. Lett.}, vol. 8, no. 5, pp. 1410-1414, Oct. 2019.

\bibitem{Wu-05} R. Long, Y.-C. Liang, Y. Pei, and E. G. Larsson, “Active reconfigurable intelligent surface aided wireless communications,” \emph{IEEE Trans. Wireless Commun.}, vol. 20, no. 8, pp. 4962–4975, Aug. 2021. 


\bibitem{Wu-06} K. Zhi, C. Pan, H. Ren, et al., ``Active RIS versus passive RIS: which is superior with the same power budget?" \emph{IEEE Wireless Commun. Letters,} vol. 26, no. 5, pp. 1150-1154, May. 2022.


\bibitem{Wu-07} M. Di Renzo, K. Ntontin, et al., ``Reconfigurable intelligent surfaces vs. relaying: Differences, similarities, and performance comparison," \emph{IEEE Open J. Commun. Soc.}, vol. 1, pp. 798–807, 2020.


\bibitem{Wu-08} B. Ning, Z. Chen, Z. Tian, et al., ``Joint power allocation and passive beamforming design for IRS-assisted physical-layer service integration," \emph{IEEE Trans. Wireless Commun.}, vol. 20, no. 11 pp. 7286-7301, Nov. 2021.


\bibitem{Wu-09} H.-M. Wang, Q.-Y. Yin, et al., ``Hybrid cooperative beamforming and jamming for physical-layer security of two-way relay networks," \emph{IEEE Trans. Inf. Forensics Secur.}, vol. 8, no. 12, pp. 2007–2020, Oct. 2013. 

\bibitem{Wu-10} S. Hong, C. Pan, H. Ren, et al., ``Robust transmission design for intelligent reflecting surface-aided secure communication systems with imperfect cascaded CSI," \emph{IEEE Trans. Wirel. Commun.}, vol. 20, no. 4, pp. 2487-2501, Apr. 2021.

\bibitem{Wu-11} H.-M. Wang, K. Huang, Q. Yang, and Z. Han, ``Joint source-relay secure precoding for MIMO relay networks with direct links," \emph{IEEE Trans. Commun.}, vol. 65, no. 7, pp. 2781-2793, Jul. 2017.






\end{thebibliography}
\end{document}